\documentclass[conference]{IEEEtran}
\IEEEoverridecommandlockouts
\usepackage{cite}
\ifCLASSINFOpdf
\usepackage[pdftex]{graphicx}
\else
\fi
\usepackage{amsmath}
\usepackage{amssymb}

\usepackage{soul}
\usepackage{amsmath}
\usepackage{tikz}
\usepackage{amssymb}

\usepackage{dirtytalk}
\usepackage{tabu}
\usepackage{comment}
\usepackage{todonotes}
\usepackage{epsf}
\usepackage{epsfig}
\usepackage{times}
\usepackage{epsfig}
\usepackage{graphicx}
\usepackage{bbold}
\usepackage{mathtools}
\usepackage{mathrsfs}
\usepackage{amssymb,amsmath}
\usepackage{pdfpages}
\usepackage{epstopdf}

\usepackage{dsfont}
\usepackage{lettrine} 

\usepackage{amsmath,epsfig,amssymb,algorithm,algpseudocode,amsthm,cite,url}

\usepackage[justification=centering]{caption}

\usepackage{subcaption}
\usepackage{bbm}
\usepackage{algorithm}% http://ctan.org/pkg/algorithm
\usepackage{algpseudocode}
\usepackage{algpseudocode}% http://ctan.org/pkg/algorithmicx
\algnewcommand{\LineComment}[1]{\State \(\triangleright\) #1}
\usepackage{csquotes}
\usepackage{amsmath,amssymb}
\DeclareFontFamily{OMX}{yhex}{}
\DeclareFontShape{OMX}{yhex}{m}{n}{<->yhcmex10}{}
\DeclareSymbolFont{yhlargesymbols}{OMX}{yhex}{m}{n}
\DeclareMathAccent{\wideparen}{\mathord}{yhlargesymbols}{"F3}
\usepackage{verbatim}
\usepackage{arcsaops}
\usepackage[english]{babel}
\usepackage{amsmath,amssymb}

\usepackage{verbatim}

\newtheorem{theorem}{\bf Theorem}
\let\oldtheorem\theorem
\renewcommand{\theorem}{\oldtheorem\normalfont}

\newtheorem{proposition}{\bf Proposition}
\let\oldproposition\proposition
\renewcommand{\proposition}{\oldproposition\normalfont}
\newtheorem{lemma}{\bf Lemma}
\let\oldlemma\lemma
\renewcommand{\lemma}{\oldlemma\normalfont}
\newtheorem{example}{Example}
\let\oldexample\example
\renewcommand{\example}{\oldexample\normalfont}

\newtheorem{definition}{\bf Definition}
\let\olddefinition\definition
\renewcommand{\definition}{\olddefinition\normalfont}
\newtheorem{remark}{Remark}
\let\oldremark\remark
\renewcommand{\remark}{\oldremark\normalfont}
\usepackage{changes}
\definechangesauthor[name={Ali Zadeh}, color=blue]{ali}
\setremarkmarkup{(#2)}

\usepackage{color}
\usepackage{xcolor}
\usepackage{dsfont}
\usepackage{subcaption}

\usepackage{tikz}
\usetikzlibrary{arrows}

\begin{document}
	%
	% paper title
	% Titles are generally capitalized except for words such as a, an, and, as,
	% at, but, by, for, in, nor, of, on, or, the, to and up, which are usually
	% not capitalized unless they are the first or last word of the title.
	% Linebreaks \\ can be used within to get better formatting as desired.
	% Do not put math or special symbols in the title.
	\title{Stochastic Optimization and Control Framework for 5G Network Slicing with Effective Isolation 
 }

	% author names and affiliations
	% use a multiple column layout for up to three different
	% affiliations
	% 	\author{\IEEEauthorblockN{Ali T. Zadeh K., Walid Saad}
	% 		\IEEEauthorblockA{Wireless@VT, Electrical and Computer Engineering Department, Virginia Tech, VA, USA,\\
	% 			Georgia Institute of Technology\\
	%           			Atlanta, Georgia 30332--0250\\
	% 			Email: http://www.michaelshell.org/contact.html}
	%		\and
	%		\IEEEauthorblockN{Merouane Debbah}
	%		\IEEEauthorblockA{Twentieth Century Fox\\
	%			Springfield, USA\\
	%			Email: homer@thesimpsons.com}
	%		\and
	%		\IEEEauthorblockN{James Kirk\\ and Montgomery Scott}
	%		\IEEEauthorblockA{Starfleet Academy\\
	%			San Francisco, California 96678--2391\\
	%			Telephone: (800) 555--1212\\
	%			Fax: (888) 555--1212}}
	
	% conference papers do not typically use \thanks and this command
	% is locked out in conference mode. If really needed, such as for
	% the acknowledgment of grants, issue a \IEEEoverridecommandlockouts
	% after \documentclass
	
	% for over three affiliations, or if they all won't fit within the width
	% of the page, use this alternative format:
	% 
    
	\author{\IEEEauthorblockN{Ali Taleb Zadeh Kasgari and
			Walid Saad} 
		\IEEEauthorblockA{Electrical and Computer Engineering Department, Blacksburg, Virginia Tech, VA, USA,
			Emails:\{alitk,walids\}@vt.edu}

	\thanks{This research was supported by the U.S. National Science Foundation under Grants CNS-1460316 and, in part, by the Office of Naval Research (ONR) under Grant N00014-15-1-2709.}	\vspace{-3em}
}
	% 	Email: homer@thesimpsons.com}
	% 	\IEEEauthorblockA{\IEEEauthorrefmark{3}Starfleet Academy, San Francisco, California 96678-2391\\
	% 	Telephone: (800) 555--1212, Fax: (888) 555--1212}
	% 	\IEEEauthorblockA{\IEEEauthorrefmark{4}Tyrell Inc., 123 Replicant Street, Los Angeles, California 90210--4321}}

	% use for special paper notices
	%\IEEEspecialpapernotice{(Invited Paper)}

	% make the title area
	\maketitle
	
	% As a general rule, do not put math, special symbols or citations
	% in the abstract
    
	\begin{abstract}
	Network slicing is an emerging technique for providing resources to diverse wireless services  with heterogeneous quality-of-service needs. However, beyond satisfying end-to-end requirements of network users, network slicing needs to also provide isolation between  slices so as to prevent one slice's faults and congestion from affecting  other slices.  In this paper,  the problem of network slicing is studied in the context of a wireless system having a time-varying number of users  that require two types of slices: reliable low latency (RLL) and self-managed (capacity limited) slices. To address this problem, a novel control framework for stochastic optimization is proposed based on the Lyapunov drift-plus-penalty method. This new framework enables the  system to  minimize power, maintain slice isolation, and provide reliable  and low latency end-to-end communication  for RLL slices.
    Simulation results show that the proposed approach can maintain the system's reliability while providing effective slice isolation in the event of sudden changes in the network environment.
    $\vspace{-.6em}$
	\end{abstract}

	% no keywords

	% For peer review papers, you can put extra information on the cover
	% page as needed:
	% \ifCLASSOPTIONpeerreview
	% \begin{center} \bfseries EDICS Category: 3-BBND \end{center}
	% \fi
	%
	% For peerreview papers, this IEEEtran command inserts a page break and
	% creates the second title. It will be ignored for other modes.
	\IEEEpeerreviewmaketitle

	\section{Introduction}
	Next-generation 5G cellular networks must support a plethora of bandwidth-intensive applications \cite{shafi2017jsac}.   Examples of such applications include ultra-low latency communication, virtual reality, unmanned aerial vehicles, and Internet of things (IoT) \cite{mozaffari2016UAV,chen2017virtual}. %\cite{ejaz2016IoT}.
    Accommodating these diverse wireless applications in 5G networks while ensuring their effective co-existence is a major challenge. To address this challenge, the concept of wireless network virtualization has recently emerged \cite{Van2017WnvSurvey}. In a virtualized wireless network, the physical wireless infrastructure and radio resources are abstracted and isolated into a number of virtual networks. This allows the network and infrastructure to be shared by several service providers, which can decrease costs, improve efficiency, and deliver the desired quality-of-service of emerging services. 
    
    Sharing network resources between these virtual networks creates the concept of \emph{network slicing}. 
	In essence, a network slice is created on top of the physical network such that it is not differentiable from a separate physical network \cite{debbah2017algorithmic}. Each slice acts as a physical dedicated network. Virtual network slices must be isolated from one another such that congestion or faults on one slice will not affect other slices. The concept of \emph{isolation} is a central goal in network slicing, as it enables network slices to function independently from each other \cite{debbah2017algorithmic}. One natural way to slice the network and provide isolation is to allocate physically separated resources to different services. For instance, allocating a different set of physical resource blocks (PRBs) to each service can facilitate  isolation between different network slices. However, this approach is inefficient because channel conditions are changing over time.
	Resource provisioning in virtual network slices is different from the conventional resource allocation problems. This is due to the fact that, in addition to minimizing power and satisfying the slices' requirements, which are the main goals of  conventional resource allocation problems, maintaining slice isolation in the network is a challenging issue in resource provisioning of the virtual network slices\cite{foukas2017slicing}.
	
	Resource allocation in wireless network virtualization has attracted significant recent attention, such as in \cite{Shi2017TVT,paraeefard2015slicing,parsaeefard2016access,kamel2014DynamicSlicing,hu2016DynamicSlicing}. For instance, in  \cite{Shi2017TVT}, the authors investigated energy efficiency and delay tradeoffs in virtual wireless networks using Lyapunov optimization and a heuristic algorithm. However, the work in \cite{Shi2017TVT} does not account for isolation between slices. The work in \cite{paraeefard2015slicing} studied the problems of resource provisioning and admission control for OFDMA-based virtualized networks. In this work, the authors considered two types of slices: rate-based slices with the minimum required rate, and resource-based slices with minimum power and sub-carrier requirement. However, allocating a minimum number of sub-carriers to the various network slices cannot guarantee any end-to-end performance for the slice tenants.
    %the service providers (SPs) do not need a minimum number of sub-carriers or power as much as they need their end-to-end requirements to be satisfied. 
     Also, the notion of slice isolation has not been studied in \cite{paraeefard2015slicing}. In \cite{parsaeefard2016access}, the authors studied the joint problem of base station assignment, sub-carrier, and power allocation. In particular, their goal was to meet the  minimum required rate per slice while maximizing the network sum rate. However, the work in \cite{parsaeefard2016access} does not take into account the effect of overloading in any of the slices. Overloading essentially occurs if the number of users in one slice increases substantially. Since slices are not isolated in \cite{parsaeefard2016access}, overloading in one slice makes the resource allocation problem infeasible.

	In \cite{kamel2014DynamicSlicing}, the authors proposed a scheme for allocating the PRBs provided by one infrastructure provider (InP) to multiple SPs. 
	The work in \cite{hu2016DynamicSlicing} introduced a novel scheme for slicing and scheduling for virtual wireless networks based on a dynamic assignment of a certain number of PRBs to each virtual network (VN) to provide service to the users. 
	Although \cite{kamel2014DynamicSlicing} and \cite{hu2016DynamicSlicing} have considered isolation in network slicing, they implement it by simply adding a constraint for the minimum number of PRBs in each slice. Such an approach can, in general, be inefficient due to the fact that it neglects the end-to-end service requirements. For instance, one slice may have a certain number of PRBs in which the users may all be experiencing a poor channel condition.
Clearly, while network slicing has been extensively studied in the literature, the existing works are not robust to overloading in slices, do not consider optimal slice isolation, and neglect end-to-end service requirement. 
%Lyapunov optimization and drift-penalty-method are proposed in \cite{Neely2008infocom} and has been used in the literature for wireless resource allocation problems \cite{ahmed2015lyapunov,ahmed2013globecom,joshi2017lyapunov}, including recently in wireless network virtualization \cite{shi2017slicingTVT}. However, it is mostly used for minimizing the average power subject to a time-average constraint. In this work, we propose a framework which is equivalent to Lyapunov optimization in a special case. Using this framework, we will be able to not only implement drift-plus-penalty, but also achieve faster and more robust stochastic optimization algorithms.  
	
	The main contribution of this paper is a novel framework for enabling network slicing with effective isolation. In particular, we formulate the joint problem of power and PRB allocation in a network with a time-varying number of users. We consider two types of slices: self-managed slices to which the InP provides a certain capacity, and reliable low latency (RLL) slices for which the InP guarantees end-to-end delay and reliability. Then, using stochastic optimization, we model the problem of minimizing power while preserving isolation of the slices from one another. Finally, we propose a control framework for stochastic optimization problems and use it to solve our network slicing problem. We then prove that our proposed framework is a generalization of the drift-plus-penalty algorithm \cite{neely2010stochastic}.
Simulation results show that our method can satisfy end-to-end requirements of slices while providing isolation between the slices. 
	
	The rest of the paper is organized as follows. Section
	\ref{sec:SysModel} introduces the system model. Sections III and IV present, respectively, the stochastic optimization model and the proposed control framework.
	Section V presents the simulation results and conclusions are drawn
	in Section VI.
	\section{System Model}\label{sec:SysModel}
	Consider the downlink of a  cellular network having a single base station (BS) with one infrastructre provider. SPs have a contract with the InP for one slice of the network with guaranteed end-to-end requirement in terms of rate, maximum tolerable delay, and reliability. The InP charges the SPs based on these three parameters. However, the number of the users for each SP should not exceed a certain threshold. This is due to the fact that the SP is responsible for the maximum number of users in each slice and, if this number exceeds the pre-set threshold, there will be a quality-of-service (QoS) drop for only the users of that particular slice. This approach will therefore provide a natural way for slice isolation.
	The total number of users in the system at time slot $t$ is $N(t)$ and the total number of PRBs is $K$. We assume that slice $s$ has $N_s(t)$  users at time slot $t$.
    The maximum arrival rate for each user in slice $s$  follows a Poisson distribution with parameter $a_s(t)$. We assume that the packet lengths for each user follow an exponential distribution. Hence, the packet queues at the BS follow an M/M/1 model.   
	
	We categorize the network slices into two types based on the application: a) \emph{Reliable low latency (RLL) slices}: these are slices with applications that require strict end-to-end QoS requirements with a certain reliability, and b) \emph{Self-managed slices}: these are slices in which only the total capacity of the slice matters to the SP. In self-managed slices, SPs purchase the capacity and they manage it themselves. Self-managed slices are typically leased to a mobile Internet provider and used for various services whose QoS mostly pertains to capacity demands. In RLL slices, InPs provide reliable, low latency  end-to-end slices to SPs. An example of applications for such slices is virtual reality.
	 
	 The network must determine the PRB and power allocation needed for network slicing in a way that the requirements for both types of slices will met while mitigating their mutual interdependence. In RLL slices, each SP that owns a certain slice $s$ has a contract for a maximum of $N_s^{\max}$ users, with a maximum tolerable delay requirement $D_s^{\max}$, and target reliability $\chi_s$. 
	 Here, reliability is essentially defined as minimum probability with which the delay is bounded by $D_s^{\max}$. 
	InP does not guarantee end-to-end reliability and delay for self-managed slices, and  the SP is responsible for using the provided capacity to ensure end-to-end QoS for its users. The InP's only responsiblity is to ensure that the capacity of these slices will not be affected by other slices, i.e., the InP is only in charge of isolation.

	The data rate for each user $i$ is given by:
	\begin{equation}
	r_i(t)=B \sum_{j=1}^{K} \rho_{ij}(t)\,\log_2\left(1+\frac{p_{ij}(t) h_{ij}(t)}{\sigma^2}\right),
	\end{equation}
	where 	$p_{ij}(t)$ is the transmit power between the BS and user $i$ over PRB $j$ at time $t$. $\sigma^2$ is noise variance, $h_{ij}(t)$ is the time-varying Rayleigh fading channel gain, and $\rho_{ij}(t)$ is an indicator function such that $\rho_{ij}(t)=1$, if PRB $j$ is allocated to user $i$ at time slot $t$; otherwise $\rho_{ij}(t) =0$.  $B$ is the bandwidth of each PRB. Given this model, our goal is to minimize the power while maintaining slice isolation and providing end-to-end QoS. Formally, this network slicing resource allocation problem is formulated as follows:
$\vspace{-.3em}$	
	\begin{subequations}\label{eq:main_problem}
		\begin{align}
		\min_{p_{ij},\rho_{ij}} \quad &\lim _{t\to \infty} \frac{1}{t} \sum_{\tau=1}^{t-1} \sum_{i=1}^{N(\tau)} \sum_{j=1}^{K} p_{ij}(\tau), \label{eq:main_problem:objective_function} \\
		\text{s.t} \quad & \lim _{t\to \infty} \frac{1}{t} \sum_{\tau=1}^{t-1}\sum_{i=1}^{N_s(\tau)} r_i(\tau) = C_s  \forall s \in \mathcal{S}_e,\label{eq:main_problem:cons_elastic}\\
		& \text{P}\{D_i>D_s^{\max}\}<1-\chi_s, \nonumber \\
		&\forall i=1,\cdots,N_s(t),\, N_s(t)<N_s^{\max},\quad \forall s \in \mathcal{S}_v, \forall t,\label{eq:main_problem:cons_end_to_end} \\
		&p_{ij}(t) \geq 0, \quad \rho_{ij}(t) \in \{0,1\}, \nonumber\\
		&\forall i =1,\cdots, N,\quad j=1,\cdots,K , \quad \forall t.
		\end{align}
	\end{subequations}
	
	$\mathcal{S}_v$ is the set of RLL slices with an end-to-end QoS contract, and $\mathcal{S}_e$ is the set of self -managed slices with only a capacity contract. $C_s$ is the  capacity limit  for self-managed slices. $N_s^{\max}$ is maximum number of users in RLL slices.
    %An example of application for RLL slices is tactile Internet, and for the self-managed slices is machine type communications.
    
    The InP seeks to allocate resources to the users in each slice so that:
	\begin{enumerate}
		\item The capacity of each self-managed slice is provided according to the contract between the SP and the InP.
		\item The end-to-end QoS requirement in RLL slices for the SP and the InP is satisfied with the predetermined reliability.
		\item Slices are isolated from each other, that is, increasing the number of users in each slice does not affect the other slices.
	\end{enumerate}
	Problem (\ref{eq:main_problem}) is a mixed-integer stochastic optimization which is an NP-hard problem. Also, since the number of users in the system changes with time, preserving isolation  in the event of a sudden change in the number of users is challenging.

	\section{Resource Allocation with Guaranteed Isolation}
	We first use a Lyapunov optimization approach \cite{neely2010stochastic} to solve problem (\ref{eq:main_problem}). In this approach, we convert constraints (\ref{eq:main_problem:cons_elastic}) and (\ref{eq:main_problem:cons_end_to_end}) into a mathematically tractable form. Subsequently, we propose a novel method to allocate resources so that the three aforementioned conditions are satisfied. 
	
	Using drift-plus-penalty method \cite{neely2010stochastic}, we first create virtual queues for (\ref{eq:main_problem:cons_elastic}). The stability of the virtual queues can satisfy (\ref{eq:main_problem:cons_elastic}). 
	To define the virtual queues, first, we rewrite (\ref{eq:main_problem:cons_elastic}) as:
	\begin{equation}
	\lim _{t\to \infty} \frac{1}{t} \sum_{\tau=1}^{t-1}\sum_{i=1}^{N_s(\tau)} r_i(\tau) - \lim _{t\to \infty} \frac{1}{t} \sum_{\tau=1}^{t-1}C_s<0.
	\end{equation}
	Then, we can define the virtual queue for self-managed slice $s$ as:
    $\vspace{-.6em}$
	\begin{equation}
    \vspace{-.3em}
	F_s(t+1)=F_s(t)+ \sum_{i=1}^{N_s(t)} r_i(t) - C_s, \,\, \forall s \in \mathcal{S}_e.
    \vspace{-.3em}
	\end{equation}
	%We can define $y_s(t)$ as
	%\begin{equation}
	%	y_s(t)=\sum_{i=1}^{N_s(t)} r_i(t) - C_s, \quad \forall s \in \mathcal{S}_e
	%\end{equation} 
	%
	% 
    
	Here, $F_s(t)$ can be either negative or positive. It is important to note that $N_s(t)$ can change in each time slot. We will show that our proposed stochastic optimization algorithm is robust to such changes. This is due to the fact that we will use the feedback from $F_s(t)$ as an input to the resource allocation system. Hence, changing $N_s(t)$ will change $F_s(t)$, and will consequently reduce the number of allocated resources to the users in slice $s$. This will allow isolation of self-managed slices $s \in \mathcal{S}_e$ from  other slices. 
	
	 Next, we must convert the RLL slices' end-to-end  QoS metrics to a physical layer metric so that we can use it in resource allocation. 
	Since we have an M/M/1 queue, we can write (\ref{eq:main_problem:cons_end_to_end}) as  \cite{Kasgari2017Asilomar}
	\begin{equation}\label{eq:converted:constrain:end_to_end}
	\lim _{t\to \infty} \frac{1}{t} \sum_{\tau=1}^{t-1} e^{(r_i(\tau)-a_i)D_s^{\max}}<1-\chi_s, \quad \forall s \in \mathcal{S}_v,
	\end{equation}
	and, then, (\ref{eq:converted:constrain:end_to_end}) can be rewritten as:
	\begin{equation}
	\lim _{t\to \infty} \frac{1}{t} \sum_{\tau=1}^{t-1} e^{(r_i(\tau)-a_i)D_s^{\max}}-\lim _{t\to \infty} \frac{1}{t} \sum_{\tau=1}^{t-1}(1-\chi_s)<0, \, \forall s \in \mathcal{S}_v.
	\end{equation}
	Hence, we can write the virtual queues for self-managed slices:
	\begin{equation}
	G_s(t+1)=\max\{G_s(t)+ y_s(t),0\}, \quad \forall s \in \mathcal{S}_v,
	\end{equation} 
	where $y_s(t)$ is defined as:
	\begin{equation}\label{eq:ys}
	y_s(t)=e^{(r_i(\tau)-a_i)D_s^{\max}}-(1-\chi_s), \quad \forall s \in \mathcal{S}_v. 
	\end{equation}

     We can see that
    \begin{align}
    &F_s(t+1)-F_s(t)=\sum_{i=1}^{N_s(t)} r_i(t) - C_s,\\
    &G_s(t+1)-G_s(t)\leq\sum_{i=1}^{N_s(t)} r_i(t) - C_s.
    \end{align}
    Then, using the telescoping rule for series, we can write:
 	\begin{align}
 		&\sum_{\tau=0}^{t-1} \sum_{i=1}^{N_s(\tau)} r_i(\tau) - C_s= \sum_{\tau=0}^{t-1} F_s(\tau+1)-\sum_{\tau=0}^{t-1} F_s(\tau)\nonumber\\
 		&=F_s(t)-F_s(0), \label{eq:init_condition_F}\\
        &\sum_{\tau=0}^{t-1} y_s(\tau)\leq \sum_{\tau=0}^{t-1} G_s(\tau+1)-\sum_{\tau=0}^{t-1} G_s(\tau)=G_s(t)-G_s(0). \label{eq:init_condition_G}
        \vspace{-.6em}
	 \end{align} 
    From (\ref{eq:init_condition_F}) and (\ref{eq:init_condition_G}), we can see that, if the initial conditions for queues $G_s(0)$ and $F_s(0)$ are finite, and 
    \begin{equation}
    \vspace{-.6em}
    \lim_{t\to \infty} \frac{1}{t} F_s(t)=\lim_{t\to \infty} \frac{1}{t} G_s(t)=0,
    \end{equation}
    i.e., the virtual queues are mean-rate stable, then constraints (\ref{eq:main_problem:cons_end_to_end}) and (\ref{eq:main_problem:cons_elastic}) are satisfied.
	 To make virtual queues $F_s(t)$ and $G_s(t)$ mean-rate stable, we define a Lyapunov function for $F_s(t)$ and $G_s(t)$, and  minimize the Lyapanuov drift for each queue. We can define the Lyapunov function for $F_s(t)$ and $G_s(t)$ as
	\begin{equation}
	L(t)=\frac{1}{2}\sum_{s\in \mathcal{S}_e} F_s(t)^2+\frac{1}{2}\sum_{s\in \mathcal{S}_v} G_s(t)^2.
	\end{equation}
	Therefore, we can find the Lyapunov drift for $F_s(t)$ and $G_s(t)$:
	\begin{align}\label{eq:main_problem:lyapunov_drift}
	\Delta &L(t)=L(t+1)-L(t)\nonumber\\
	&=\frac{1}{2}\sum_{s\in \mathcal{S}_e}\big\{\max\{F_s(t)+ \sum_{i=1}^{N_s(t)} r_i(t) - C_s,0\}^2-F_s(t)^2\big\}\nonumber\\
	&+\frac{1}{2}\sum_{s\in \mathcal{S}_v}\big\{\max\{G_s(t)+ y_s(t),0\}^2-G_s(t)^2\big\}\nonumber\\
	&\leq \frac{1}{2}\sum_{s\in \mathcal{S}_e} \left(\sum_{i=1}^{N_s(t)} r_i(t) - C_s\right)^2+\frac{1}{2}\sum_{s\in \mathcal{S}_v}y_s(t)^2\nonumber\\
	&+\sum_{s\in \mathcal{S}_e}\left(\sum_{i=1}^{N_s(t)} r_i(t) - C_s\right)F_s(t)+  \sum_{s\in \mathcal{S}_v} y_s(t)G_s(t).
	\end{align}

	%where $I_{\mathcal{S}_e}$ and $I_{\mathcal{S}_v}$ are indicator functions, i.e.,
	%\begin{equation}
	%	I_{\mathcal{S}_e}=
	%	\begin{cases}
	%		1, &s\in \mathcal{S}_e\\
	%		0, & \text{o.w.}
	%	\end{cases}
	%\end{equation}
	
	If we assume that the first term in the right hand side of (\ref{eq:main_problem:lyapunov_drift}) is bounded, i.e.,
	\begin{equation}
	\frac{1}{2}\sum_{s\in \mathcal{S}_e} \left(\sum_{i=1}^{N_s(t)} r_i(t) - C_s\right)^2+\frac{1}{2}\sum_{s\in \mathcal{S}_v}y_s(t)^2<B,
	\end{equation}
	
	then, we can write drift-plus-penalty as:
	\begin{align}
	&\Delta L(t)+\sum_{i=1}^{N(t)} \sum_{j=1}^{K} p_{ij}(t)\leq B
	+\sum_{i=1}^{N(t)} \sum_{j=1}^{K} p_{ij}(t)\nonumber \\&+\sum_{s\in \mathcal{S}_e}\left(\sum_{i=1}^{N_s(t)} r_i(t) - C_s\right)F_s(t)+  \sum_{s\in \mathcal{S}_v} y_s(t)G_s(t).
	\end{align}
	
	Now, we can transform stochastic optimization problem (\ref{eq:main_problem}) into the following optimization problem which needs to be solved by the network in each time slot:
	\begin{subequations}\label{eq:converted_problem}
		\begin{align}
		\min_{p_{ij}(t),\rho_{ij}(t)} \qquad &\sum_{i=1}^{N(t)} \sum_{j=1}^{K} p_{ij}(t)+\sum_{s\in \mathcal{S}_e}\Big(\sum_{i=1}^{N_s(t)} r_i(t) - C_s\Big)F_s(t)\nonumber\\ &+  \sum_{s\in \mathcal{S}_v} y_s(t)G_s(t), \label{eq:converted_problem:objective_function} \\
		\text{s.t} \qquad &p_{ij}(t) \geq 0, \quad \rho_{ij}(t) \in \{0,1\} \nonumber\\
		&\forall i =1,\cdots, N,\quad j=1,\cdots,K \label{eq:constrain3}, \quad \forall t.
		\end{align}
	\end{subequations}
	
\begin{figure}[!t]
		\centering
		\includegraphics[clip, trim=0cm 0cm 0cm 0cm, width=0.5\textwidth]{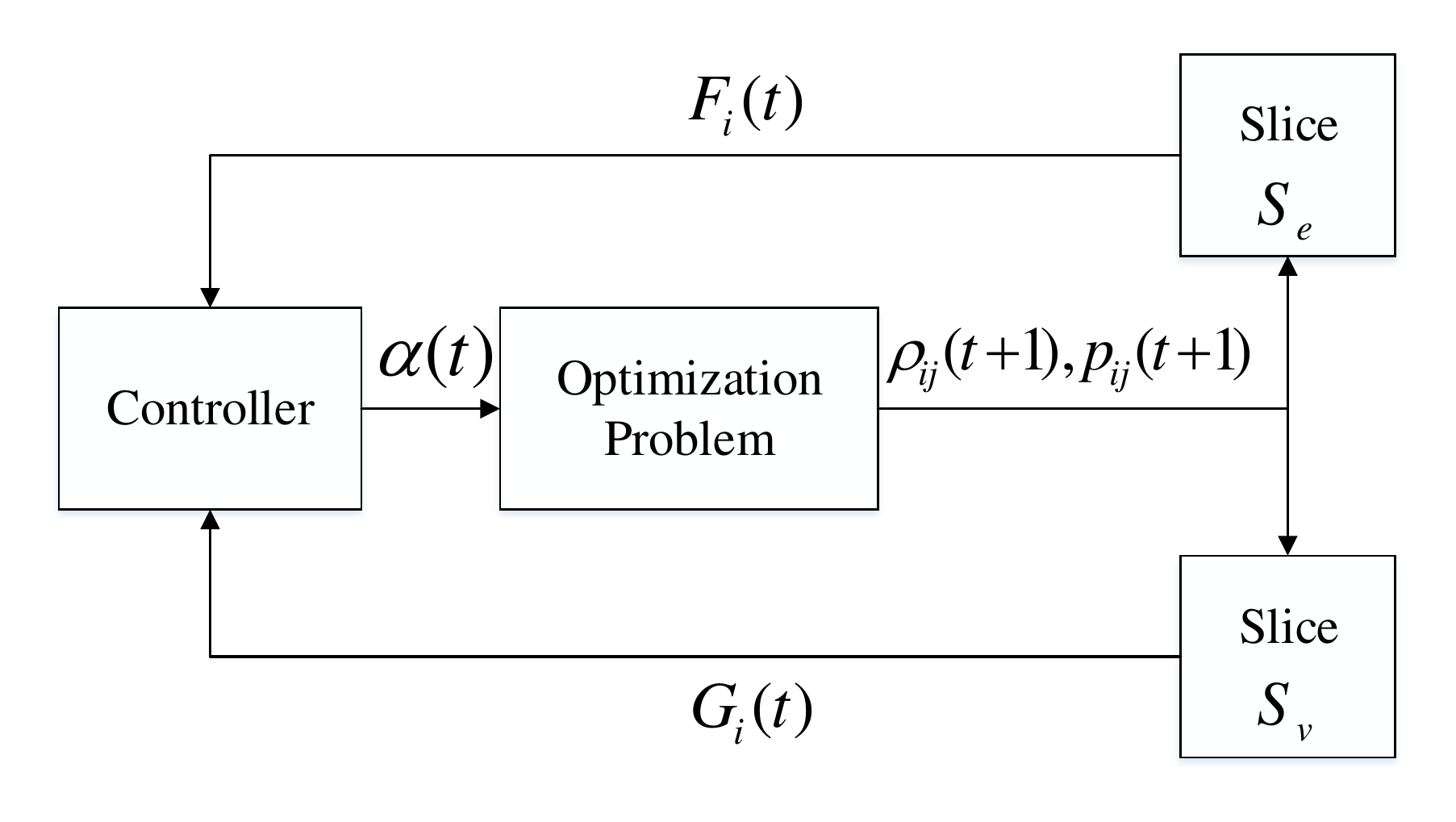}
		\caption{ٍBlock diagram of the proposed Lyapunov optimization network slicing.}
		\label{fig:lyap}
        \vspace{-.6em}
	\end{figure}

	Lyapunov optimization changes the original problem to (\ref{eq:converted_problem}). However, solving (\ref{eq:converted_problem}) greedily in each time slot has a high complexity and, hence, will not be scalable for large networks. To overcome this scalability challenge, we  propose a novel control method that can reduce the complexity of (\ref{eq:converted_problem}) by using a feedback mechanism in a decomposable optimization problem.
	The block diagram of the proposed Lyapunov optimization for the resource allocation problem is shown in Fig. \ref{fig:lyap}.
	
	\section{Stochastic Optimization Control Framework}
    Next, we propose an optimization problem that is equivalent to the Lyapunov optimization problem in (\ref{eq:converted_problem}) and we show its equivalence to the drift-plus-penalty algorithm.
	Consider the optimization problem:
	\begin{subequations}\label{eq:q:controlled problem}
		\begin{align}
		\min_{p_{ij}(t),\rho_{ij}(t)} \qquad &\sum_{i=1}^{N(t)} \sum_{j=1}^{K} p_{ij}(t),\\
		\text{s.t} \qquad &r_i(t)=\alpha_i,\\
		&p_{ij}(t) \geq 0, \quad \rho_{ij}(t) \in \{0,1\}, \nonumber\\
		&\forall i =1,\cdots, N,\quad j=1,\cdots,K \label{eq:constrain3}, \quad \forall t.
		\end{align}
	\end{subequations}
	
	Here, $\boldsymbol\alpha=\begin{bmatrix}
	\alpha_1 &\alpha_2 &\cdots &\alpha_{N(t)}
	\end{bmatrix}^T$ is the input parameter vector that controls the solution of this optimization problem. Next, we show the equivalence result.
    
    	\begin{theorem}\label{th:1}
	The optimization problem (\ref{eq:q:controlled problem}) is equivalent to the drift-plus-penalty problem in (\ref{eq:converted_problem}).
	\end{theorem}
    \begin{IEEEproof}
    If we write the Lagrangian for problem (\ref{eq:q:controlled problem}) and keep the feasibility conditions, we have
	
	\begin{subequations}\label{eq:q:controlled problem:Lagrangian}
		\begin{align}
		\max_{\lambda_i}\min_{p_{ij}(t),\rho_{ij}(t)} \qquad &\sum_{i=1}^{N(t)} \sum_{j=1}^{K} p_{ij}(t)+\sum_{i=1}^{N}\lambda_i (\alpha_i-r_i),\\
		\text{s.t} \qquad 	&p_{ij}(t) \geq 0, \quad \rho_{ij}(t) \in \{0,1\} \nonumber\\
		&\forall i =1,\cdots, N,\quad j=1,\cdots,K \label{eq:constrain3}, \quad \forall t.
		\end{align}
	\end{subequations}
	Furthermore, (\ref{eq:converted_problem:objective_function})  can be rewritten as:
	\begin{align}\label{eq:lyap_opt_simplified}
	\min_{p_{ij}(t),\rho_{ij}(t)} \qquad &\sum_{i=1}^{N(t)} \sum_{j=1}^{K} p_{ij}(t)+\sum_{s\in \mathcal{S}_e}\big(\sum_{i=1}^{N_s(t)} r_i(t)\big)F_s(t)\nonumber\\
	&+  \sum_{s\in \mathcal{S}_v} -b_s(t)\,G_s(t)\,r_s(t),
	\end{align}
	in which we have omitted $C_s F_s(t)$, because it does not contain any of the decision variables $p_{ij}(t)$ and $\rho_{ij}(t)$.
	Also, we substituted $y_s(t)$ with $-b_s(t) r_s(t)$, which is the first-order approximation of $y_s(t)$ in (\ref{eq:ys}).
	
	In problem (\ref{eq:q:controlled problem:Lagrangian}), increasing $\alpha_i(t)$ will increase $\lambda_i(t)$, and this will, in turn, change  the problem (\ref{eq:q:controlled problem:Lagrangian}) to
	\begin{equation}\label{eq:lagrange_simplified}
	\min_{p_{ij}(t),\rho_{ij}(t)} \qquad \sum_{i=1}^{N(t)} \sum_{j=1}^{K} p_{ij}(t)-\sum_{i=1}^{N}\lambda_i^* r_i,
	\end{equation}
	Where $\lambda_i^*$ is the optimal $\lambda$ for the problem (\ref{eq:q:controlled problem:Lagrangian}).
	As we can see, problems (\ref{eq:lyap_opt_simplified}) and (\ref{eq:lagrange_simplified}) are equivalent, and
	\begin{equation}\label{eq:lambda_equivalent}
	\lambda_i^*=\begin{cases}
	-F_i(t), &i\in \mathcal{S}_e,\\
	b_i(t) G_i(t), &i\in \mathcal{S}_v.
	\end{cases}
	\end{equation} 
    \end{IEEEproof}
	From Theorem \ref{th:1}, we can see that changing $\alpha_i(t)$ has the same effect as changing $F_i(t)$ and $G_i(t)$ in  (\ref{eq:lyap_opt_simplified}), i.e., using a specific type of controller which generates control signal $\boldsymbol\alpha$ such that (\ref{eq:lambda_equivalent}) is satisfied, the solution will be the same as the drift-plus-penalty algorithm. Nonetheless,  we can also observe that the drift-plus penalty is a special case of our general framework.

	\begin{figure*}[!t]
		\centering
		\includegraphics[clip, trim=0cm 0cm 0cm 0cm, width=\textwidth]{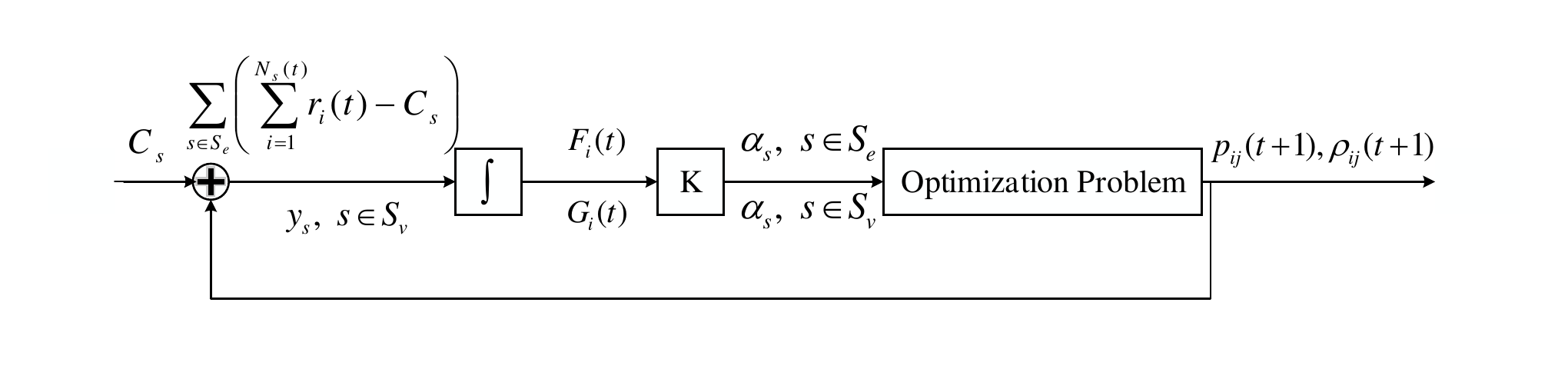}
        \vspace{-3em}
		\caption{Block diagram of the equivalent control system.}
		\label{fig:control block}
        \vspace{-2em}
	\end{figure*}
	With that in mind, the drift-plus-penalty algorithm can be modeled as a control system as shown in Fig. \ref{fig:control block}. Since the virtual queues $Q_i(t)$ and $F_i(t)$ accumulate errors, they are shown as integrators in Fig. \ref{fig:control block}. Therefore, we can see that the drift-plus-penalty algorithm consists of a proportional-integral (PI) controller. As we will see in Section {\ref{sec:simul}, this method suffers from the same drawbacks of PI controllers, that is, it is not sensitive to steady state error as long as the average error is zero. By adding another type of controller to the drift-plus-penalty method, instead of PI controller in Fig. \ref{fig:control block}, we can improve both steady state error and transient time response of the drift-plus-penalty method.

$\vspace{-1.7em}$
	\section{Simulation Results and Analysis}\label{sec:simul}
	For our simulation, we consider a network with a cell radius of $1.5$~km having four slices, two of which are self-managed slices ($s_e^1,s_e^2$), and the other two are RLL slices ($s_v^1,s_v^2$). We set the bandwidth to $B=10$~MHz. We also set $a_s(t)=1$~Mbps for $s_v^1,s_v^2$, and $\sigma^2=-173.9$~dBm. For $s_v^1$ and $s_v^2$,  $D_s^{\max}$ and $\chi$ are respectively set to $10$, $20$~ms and $99\%$, $95\%$.   We set the path loss exponent to $3$ (urban area), and the carrier frequency to $900$~MHz. The packet length for RLL slices is an exponential random variable with an average size of $10$~kbits. 
	
	\begin{figure}[!t]
		\centering
		\includegraphics[scale=.6,trim={0.6cm 0 0 0}]{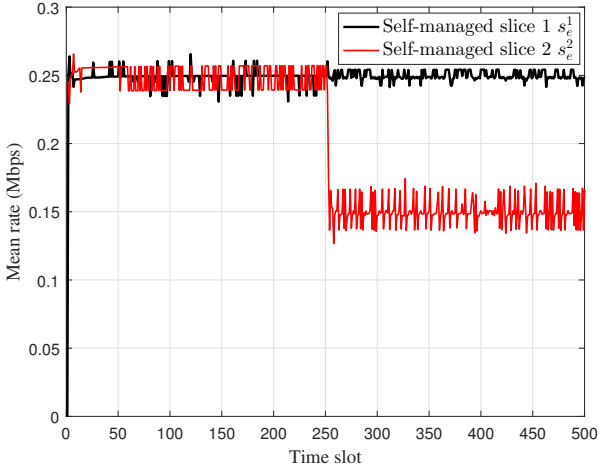}
		\caption{Mean rate of self-managed slices slices' users.}
		\label{fig:mean_rate_se}
	\end{figure}

\begin{table}[t!]
	\centering
	\begin{tabular}{ |c|c|c| } 
		\hline
		slice & \# of active users in $t<250$ & \# of active users in $t>250$ \\
		\hline
		$s_v^1$ & $1$ & $1$ \\ 
		$s_v^2$ &$1$ & $1$ \\ 
		$s_e^1$ & $5$ & $5$ \\ 
		$s_e^2$& $2$ & $5$ \\ 
		\hline
	\end{tabular}
	\caption{Specifications of the various slices.}
	\label{table:num_user}
\end{table}

\begin{table}[t!]
	\centering
	\begin{tabular}{ |c|c| } 
		\hline
		slice  &capacity\\
		\hline
		$s_v^1$  &$\text{P}(D>10\times 10^{-3})<0.01$ \\ 
		$s_v^2$ &$\text{P}(D>20\times 10^{-3})<0.05$\\ 
		$s_e^1$ &$2.5$~Mbps\\ 
		$s_e^2$ &$750$~kbps\\ 
		\hline
	\end{tabular}
	\caption{Slices capacity and end-to-end requirements.}
	\label{table:num_user}
    \vspace{-.6em}
\end{table}
The number of users in each slice is shown in Table \ref{table:num_user}.
	The total running time is $500$ time slots. At time slot $250$, we simulate a sudden increase in the number of the users in slice $s_e^2$. The number of users increases from 2 to 5 users.

	In order to evaluate slice isolation, in Fig. \ref{fig:mean_rate_se}, we show the mean rate of the users in self-managed slices. From this figure, we can see that, at the time of the sudden change in the number of users in  slice 2, the mean rate for each user in this slice decreases. However, the mean rate for the users in slice 1 does not change. This demonstrates the complete isolation of slices.

	\begin{figure}[!t]
		\centering
		\includegraphics[scale=.6,trim={0.6cm 0 0 0}]{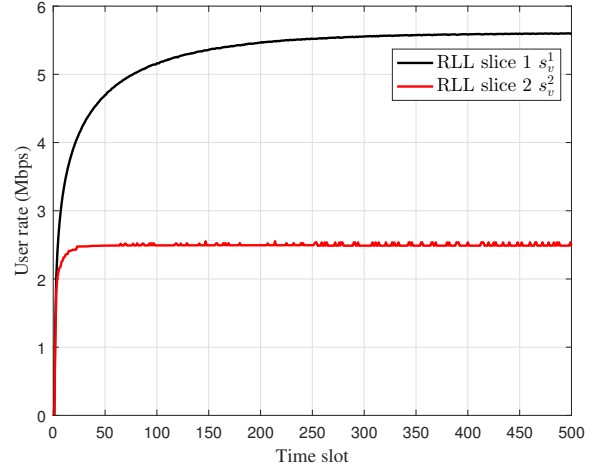}
		\caption{Rate for RLL slices.}
		\label{fig:user_rate_sv}
        \vspace{-1.6em}
	\end{figure}
	
	Fig. \ref{fig:user_rate_sv} shows the rate of RLL slices, as time evolves. As we can see, since $s_v^1$ has a strict end-to-end QoS requirement, it takes longer for the proposed algorithm to converge. From Fig. \ref{fig:user_rate_sv}, we can also see that the sudden changes in the self-managed slice does not have any effect on RLL slices. This result further confirms that our method provides complete isolation between the slices.

	\begin{figure}[!t]
		\centering
		\includegraphics[scale=.6,trim={0.6cm 0 0 0}]{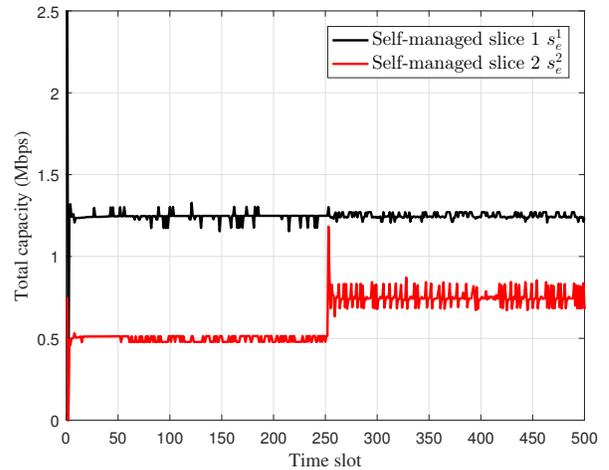}
		\caption{Total capacity that for self-managed slices.}
		\label{total_capacity_se}
        \vspace{-1.4em}
	\end{figure}
	
	Fig. \ref{total_capacity_se} shows the total capacity used by self-managed slices. The total capacity of slice $s_v^1$ is the sum of its users' arrival rates which is $1.25$~Mbps. Since slice $s_v^2$ exceeds the allocated capacity, the total capacity will be set to the determined limit which is $750$~kbps. As we can see from Fig. \ref{total_capacity_se}, there is a high overshoot in the total capacity of $s_v^2$. However, since the duration of this overshoot is very short, it does not affect the other slices. %Although this overshoot did not affect other slices, in a more crowded network, this may decrease other slices capacity. 

	\begin{figure}[!t]
		\centering
		\includegraphics[scale=.6,trim={0.6cm 0 0 0}]{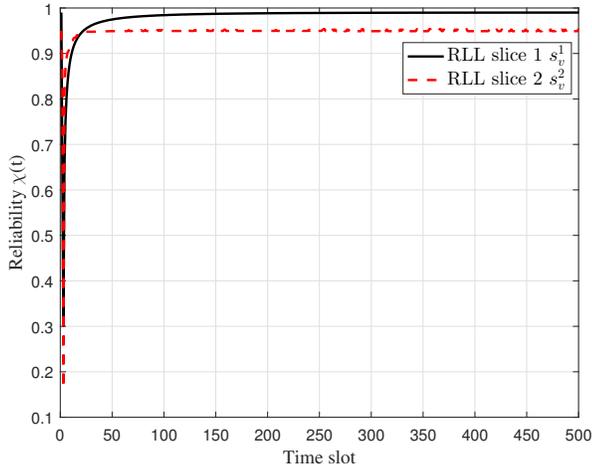}
		\caption{Reliability over time for RLL slices.}
		\label{fig:ys_sv}
        \vspace{-1.5em}
	\end{figure}

	In Fig. \ref{fig:ys_sv}, we evaluate the reliability for RLL slices. From this figure, we can first see that the reliability increases with time. Due to its strict QoS requirement, the reliability for $s_v^1$ increases slowly with time.
	
	\begin{figure}[!t]
		\centering
		\includegraphics[scale=.6,trim={0.6cm 0 0 0}]{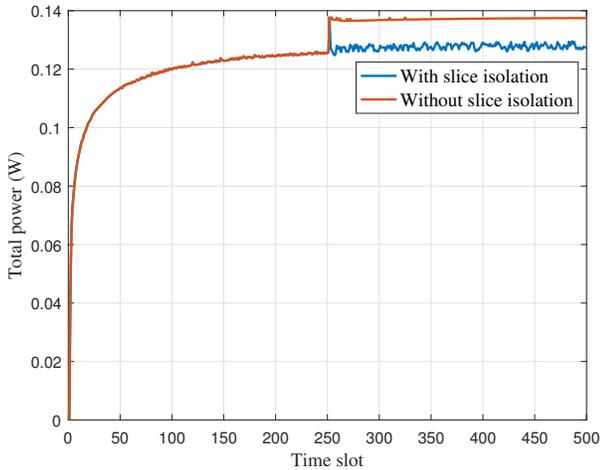}
		\caption{Total BS transmit power.}
		\label{fig:power versus delay}
        \vspace{-1.5em}
	\end{figure}
	
	Fig. \ref{fig:power versus delay} shows the total BS power resulting from our proposed algorithm and also in the case of network slicing without isolation. From Fig. \ref{fig:power versus delay}, we can see that, at the time of the sudden change in slice $s_e^2$, there is an overshoot in the power usage, which is caused by the overshoot that happened in the capacity of slice $s_e^2$. However, in our algorithm, the overshoot disappeared  quickly while in the case of no isolation, this change in the number of users will affect the total power usage in the system. This additional $8\%$ power consumption can affect the total network performance since the total power might be limited and should not be affected by an overload in one slice.
		$\vspace{-.5em}$
	\section{Conclusion}
      In this paper, we have studied the problem of network slicing with a variable number of users for two types of slices: RLL slices, with  strict latency and reliability requirements and self-managed slices with capacity requirements. We have proposed a new control framework that can solve the stochastic optimization problem of PRB and power allocation in LTE downlink in polynomial time.  We have shown that the proposed control algorithm
 minimizes power while satisfying the slices' QoS requirements and preserving complete isolation among slices.
%    have solved the problem of power minimization with complete isolation of the slices. We have proposed a new control framework for stochastic optimization problem of PRB and power allocation in LTE downlink, and  overcome complexity problem of mixed integer stochastic optimization.
    The results have shown that isolation can be provided even in the event of sudden changes in the number of users in the system. 
 $\vspace{-.3em}$
	
	%, spectrum should be  The rate that is allocated to that capacity region for each user is

	%\section{Conclusion}
	%The conclusion goes here.

	$\vspace{-1.5em}$
	
	\nocite{*}
	\bibliographystyle{IEEEtran}
	\bibliography{PaperBib.bib}

	% that's all folks
\end{document}